\documentclass{emulateapj}
\usepackage{natbib}
\usepackage{environ}

\def\apjl{{ApJ}}%
\def\apjs{{ApJS}}%
\def\aap{{A\&A}}%
\newcommand\hst{{\it HST}}
\newcommand{\kms}{\,km\,s$^{-1}$}
\newcommand{\mstellar}{\ensuremath{M_{\mathrm{stellar}}}}
\newcommand{\omatter}{\ensuremath{\Omega_{\mathrm{M}}}}

\shorttitle{The host environment of SSS17a}
\shortauthors{Pan et~al.}

\begin{document}

\title{The Old Host-Galaxy Environment of SSS17a, the First Electromagnetic Counterpart to a Gravitational Wave Source{*}}

\altaffiltext{*}{This paper includes data gathered with the 6.5 meter
  Magellan Telescopes located at Las Campanas Observatory, Chile.}

\author{
{Y.-C.~Pan}\altaffilmark{1},
{C.~D.~Kilpatrick}\altaffilmark{1},
{J.~D.~Simon}\altaffilmark{2},
{E.~Xhakaj}\altaffilmark{1},
{K.~Boutsia}\altaffilmark{3},
{D.~A.~Coulter}\altaffilmark{1},
{M.~R.~Drout}\altaffilmark{2,4},
{R.~J.~Foley}\altaffilmark{1},
{D.~Kasen}\altaffilmark{5,6},
{N.~Morrell}\altaffilmark{3},
{A.~Murguia-Berthier}\altaffilmark{1},
{D.~Osip}\altaffilmark{3},
{A.~L.~Piro}\altaffilmark{2},
{J.~X.~Prochaska}\altaffilmark{1},
{E.~Ramirez-Ruiz}\altaffilmark{1,7},
{A.~Rest}\altaffilmark{8,9},
{C.~Rojas-Bravo}\altaffilmark{1},
{B.~J.~Shappee}\altaffilmark{2,10,11}, and
{M.~R.~Siebert}\altaffilmark{1}
}

\altaffiltext{1}{
Department of Astronomy and Astrophysics, University of California, Santa Cruz, CA 95064, USA
}
\altaffiltext{2}{
The Observatories of the Carnegie Institution for Science, 813 Santa Barbara Street, Pasadena, CA 91101
}
\altaffiltext{3}{
Las Campanas Observatory, Carnegie Observatories, Casilla 601, La Serena, Chile
}
\altaffiltext{4}{
Hubble and Carnegie-Dunlap Fellow
}
\altaffiltext{5}{
Nuclear Science Division, Lawrence Berkeley National Laboratory, Berkeley, CA 94720, USA
}
\altaffiltext{6}{
Departments of Physics and Astronomy, University of California, Berkeley, CA 94720, USA
}
\altaffiltext{7}{
DARK, Niels Bohr Institute, University of Copenhagen, Blegdamsvej 17, 2100 Copenhagen, Denmark
}
\altaffiltext{8}{
Space Telescope Science Institute, 3700 San Martin Drive, Baltimore, MD 21218, USA
}
\altaffiltext{9}{
Department of Physics and Astronomy, The Johns Hopkins University, 3400 North Charles Street, Baltimore, MD 21218, USA}
\altaffiltext{10}{
Hubble and Carnegie-Princeton Fellow}
\altaffiltext{11}{
Institute for Astronomy, University of Hawaii, 2680 Woodlawn Drive, Honolulu, HI 96822, USA}

\begin{abstract}
We present an analysis of the host-galaxy environment of Swope Supernova Survey 2017a (SSS17a), the discovery of an electromagnetic counterpart to a gravitational wave source, GW170817.  SSS17a occurred 1.9~kpc (in projection; $10\farcs2$) from the nucleus of NGC~4993, an S0 galaxy at a distance of 40~Mpc.  We present a {\it Hubble Space Telescope} ({\it HST}) pre-trigger image of NGC~4993, Magellan optical spectroscopy of the nucleus of NGC~4993 and the location of SSS17a, and broad-band UV through IR photometry of NGC~4993.  The spectrum and broad-band spectral-energy distribution indicate that NGC~4993 has a stellar mass of $\log (M/M_{\sun}) = 10.49^{+0.08}_{-0.20}$ and star formation rate of 0.003\,M$_{\sun}$\,yr$^{-1}$, and the progenitor system of SSS17a likely had an age of $>$2.8~Gyr.  There is no counterpart at the position of SSS17a in the {\it HST} pre-trigger image, indicating that the progenitor system had an absolute magnitude $M_{V} > -5.8$~mag. We detect dust lanes extending out to almost the position of SSS17a and $>$100 likely globular clusters associated with NGC~4993.  The offset of SSS17a is similar to many short gamma-ray burst offsets, and its progenitor system was likely bound to NGC~4993.  The environment of SSS17a is consistent with an old progenitor system such as a binary neutron star system.
\end{abstract}


\section{Introduction}
\label{sec:introduction}
On 2017 August 17 (UT), the Laser Interferometer Gravitational Wave Observatory (LIGO) and Virgo interferometer detected a gravitational wave source from a binary neutron star (BNS) merger, GW170817 \citep[][; LIGO Scientific Collaboration and Virgo Collaboration, in preparation]{GCN21509}. Two seconds after the LIGO/Virgo detection, the Fermi Gamma-ray Space Telescope and INTErnational Gamma-Ray Astrophysics Laboratory (INTEGRAL) detected a short-duration gamma-ray burst \citep[sGRB;][]{GCN21505,GCN21507}. About 11 hours after the LIGO/Virgo trigger, our team discovered an optical transient in NGC~4993 coincident with GW170817, called Swope Supernova Survey 2017a (SSS17a; \citealt{GCN21529,Coulter17}).

SSS17a is the first detection of an electromagnetic counterpart to a gravitational wave source. This discovery marks a milestone and opens a new era in modern astronomy. The gravitational wave data suggests that SSS17a is a BNS merger, the most popular progenitor model of sGRBs \citep[e.g.,][]{1989Natur.340..126E,2014ARA&A..52...43B,2007NJPh....9...17L}. 

The host environments of astrophysical transients have long been a profitable route to understanding the nature of their progenitor systems and placing broad constraints on their properties. For example, the long-duration GRBs and sGRBs have very different host environments. While long GRBs predominantly occur in star-forming galaxies \citep[e.g.,][]{2002AJ....123.1111B}, sGRBs can be found in both star-forming and early-type galaxies \citep{2006ApJ...642..989P, 2013ApJ...769...56F}, indicating an older population. In addition, sGRBs tend to be found in more massive galaxies and generally show larger offsets from their hosts than long GRBs do \citep{2007ApJ...665.1220Z,2014ApJ...792..123B}. The distinct host properties suggest they are likely to arise from different progenitor populations.

In this work, we investigate the host environment of SSS17a, both globally and locally. By comparing our results to those from different kinds of astrophysical transients, we constrain the nature of the progenitor system.

A plan of the paper follows. In Section~\ref{sec:obs}, we describe the observations and data reduction, and Section~\ref{sec:analysis} discusses the methods used to analyze the data and show the determined host properties. The discussion and conclusions are presented in Section~\ref{sec:discussion} and \ref{sec:conclusions}, respectively. Throughout this paper, we assume $\mathrm{H_0}=70$\,km\,s$^{-1}$\,Mpc$^{-1}$ and a flat universe with $\omatter=0.3$ when necessary.

\begin{figure*}
	\centering
	\begin{tabular}{c}
		\includegraphics*[scale=0.5]{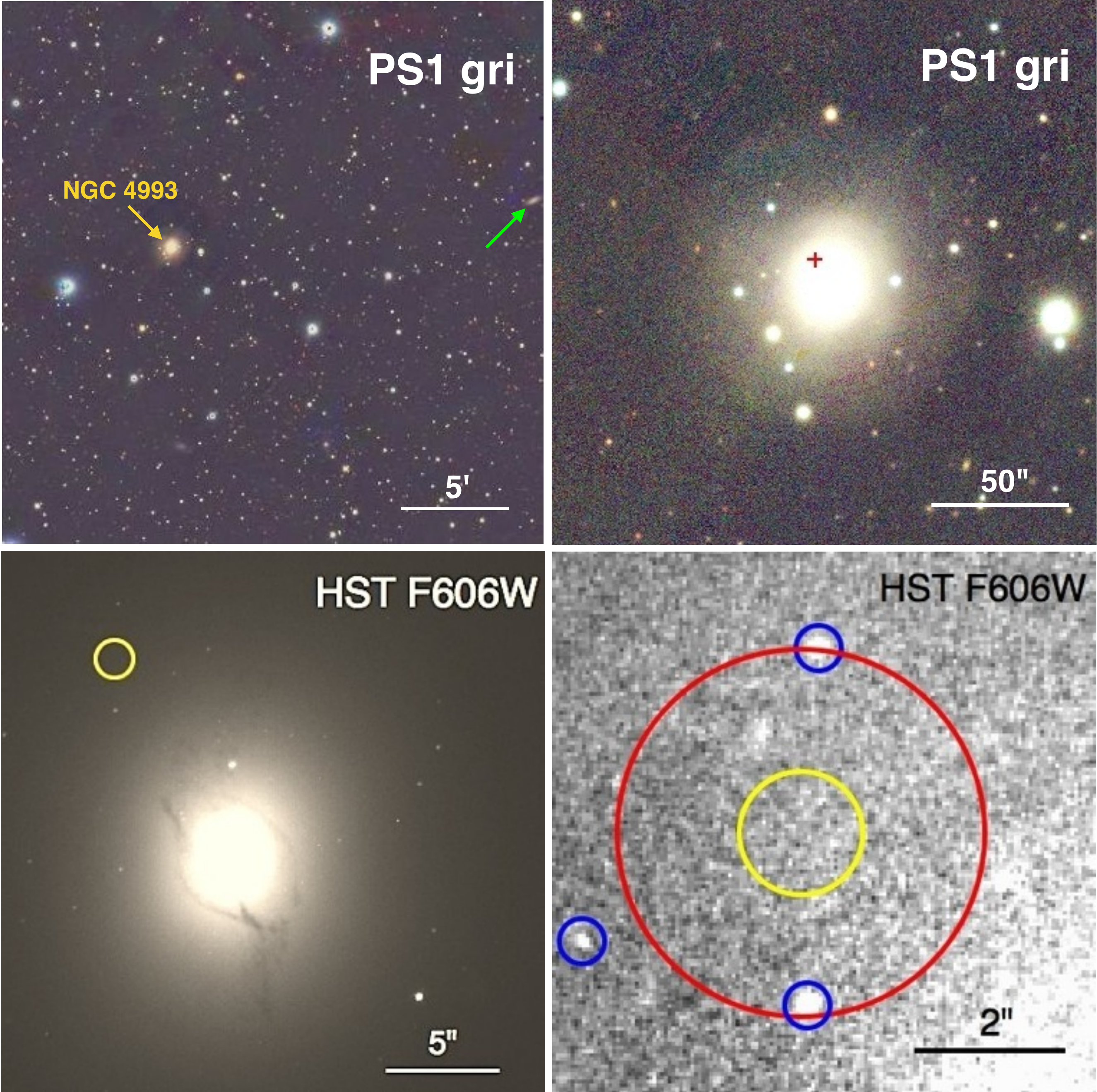}
	\end{tabular}
        \caption{{\it Upper left}: False-color (RGB channels corresponding to PS1 $irg$ filters) image of the field surrounding NGC~4993 (indicated by the yellow arrow). The green arrow marks the closest galaxy to NGC~4993 in the same galaxy group, ESO~508-G018. {\it Upper right}: Zoom-in of the upper-left image centered on NGC~4993. The red cross marks the location of SSS17a. {\it Lower left}: {\it HST}/ACS F606W image near the center of NGC~4993. The yellow circle represents the 1-$\sigma$ error circle of the SSS17a location. The dust lanes surrounding the galactic center are obvious. {\it Lower right}: Zoom-in of the lower-left image after subtracting a \textsc{galfit} model, centered on the SSS17a location. The yellow circle and red circle represent the 1-$\sigma$ and 3-$\sigma$ astrometric uncertainty, respectively. Blue circles mark the potential globular clusters near the transient location. For all images, north is up and east is left.}\label{host-field}
\end{figure*}

\section{Observations and Data Reduction}
\label{sec:obs}

SSS17a was discovered $5\farcs3$~E and $8\farcs7$~N of NGC~4993 \citep{GCN21529, Coulter17}, an early-type S0 galaxy with redshift $z = 0.009727 \pm 0.000050$ \citep{1991rc3..book.....D} in a galaxy group \citep{2011MNRAS.412.2498M}.  The transient is only 1.9\,kpc offset (projected) from NGC~4993, assuming the distance to NGC~4993 of 39.5\,Mpc based on the Tully-Fisher method \citep{2001ApJ...553...47F}.

NGC~4993 was observed by the {\it Hubble Space Telescope} ({\it HST}) with the Advanced Camera for Surveys (ACS) on April 28, 2017 (UT) in the F606W filter as part of the ``Schedule Gap Pilot'' program (Program 14840; PI Bellini).  We obtained the {\it HST} images from the Mikulski Archive for Space Telescopes (MAST). We reduced the {\it HST} image using the \textsc{drizzlepac} pipeline \citep{2015ASPC..495..281A}.  The calibrated frames were further corrected for geometric distortion, sky background, cosmic-rays and combined with \textsc{astrodrizzle}.  We registered the final, combined images using \textsc{TweakReg}.

We performed photometry on the combined \hst/ACS image following standard procedures with \textsc{dolphot}\footnote{\url{http://americano.dolphinsim.com/dolphot/}}.  The \textsc{dolphot} photometry was calibrated using the ACS/WFC $F606W$ zero point for April 28, 2017 from the ACS zero point calculator\footnote{https://acszeropoints.stsci.edu/}.

We obtained Pan-STARRS1 (PS1) $griz$ imaging of NGC~4993 from the PS1 image cutout server\footnote{http://ps1images.stsci.edu/cgi-bin/ps1cutouts} \citep{Chambers16}.  These data had been calibrated to the PS1 system following procedures described in \citet{Magnier16}.  

To measure the photometry of NGC~4993, we fit an elliptical isophote to the galaxy profile using the \textsc{IRAF} package \textsc{isophote}. We measured an {\it HST}/ACS $F606W$ AB magnitude of $12.23 \pm 0.01$\,mag. Using the same method, we measured PS1 $griz$ AB magnitudes of $12.45 \pm 0.02$, $12.14 \pm 0.02$, $11.78 \pm 0.02$, and $12.62 \pm 0.02$\,mag, respectively. In addition, we obtained far-UV (FUV) and near-UV (NUV) photometry from the Galaxy Evolution Explorer \citep[GALEX;][]{2017ApJS..230...24B}, $J\!H\!K_{s}$ near-infrared (NIR) photometry from the Two Micron All-Sky Survey \citep[2MASS;][]{2006AJ....131.1163S} and $3.6-22$~$\mu$m IR photometry from the Wide-field Infrared Survey Explorer \citep[WISE;][]{2010AJ....140.1868W}.

We examined the position of SSS17a in the {\it HST}/ACS $F606W$ image and did not detect any sources at the transient location.  Placing artificial stars on similar surface-brightness areas, we determined an AB magnitude limit at the position of SSS17a of $m_{V} > 27.2$\,mag, corresponding to $M_{V} > -5.8$\,mag at the distance of NGC~4993, consistent with limits initially reported by \citet{GCN21536}.

\begin{figure}
	\centering
	\begin{tabular}{c}
	    \includegraphics*[scale=0.51]{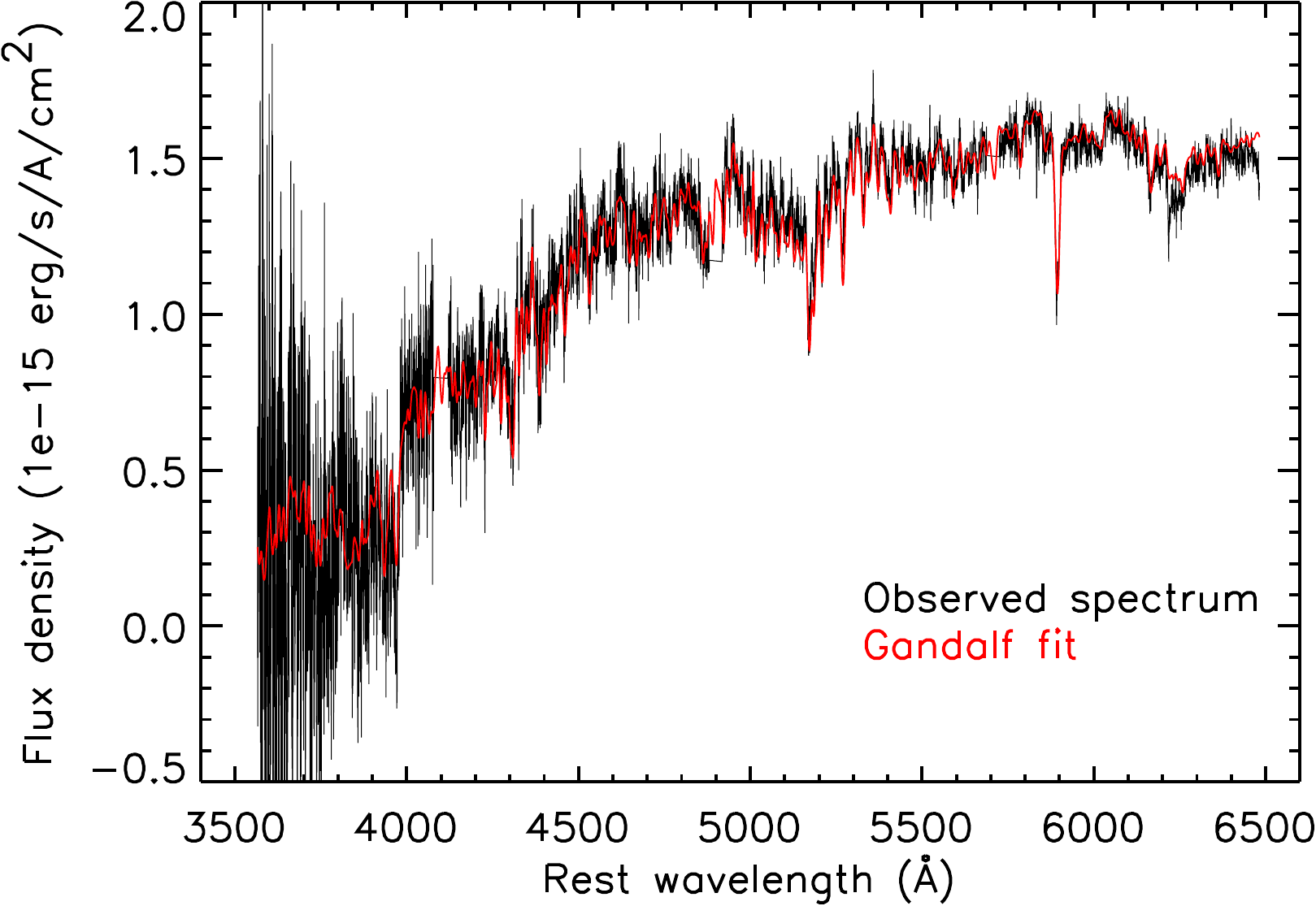}\\
		\includegraphics*[scale=0.51]{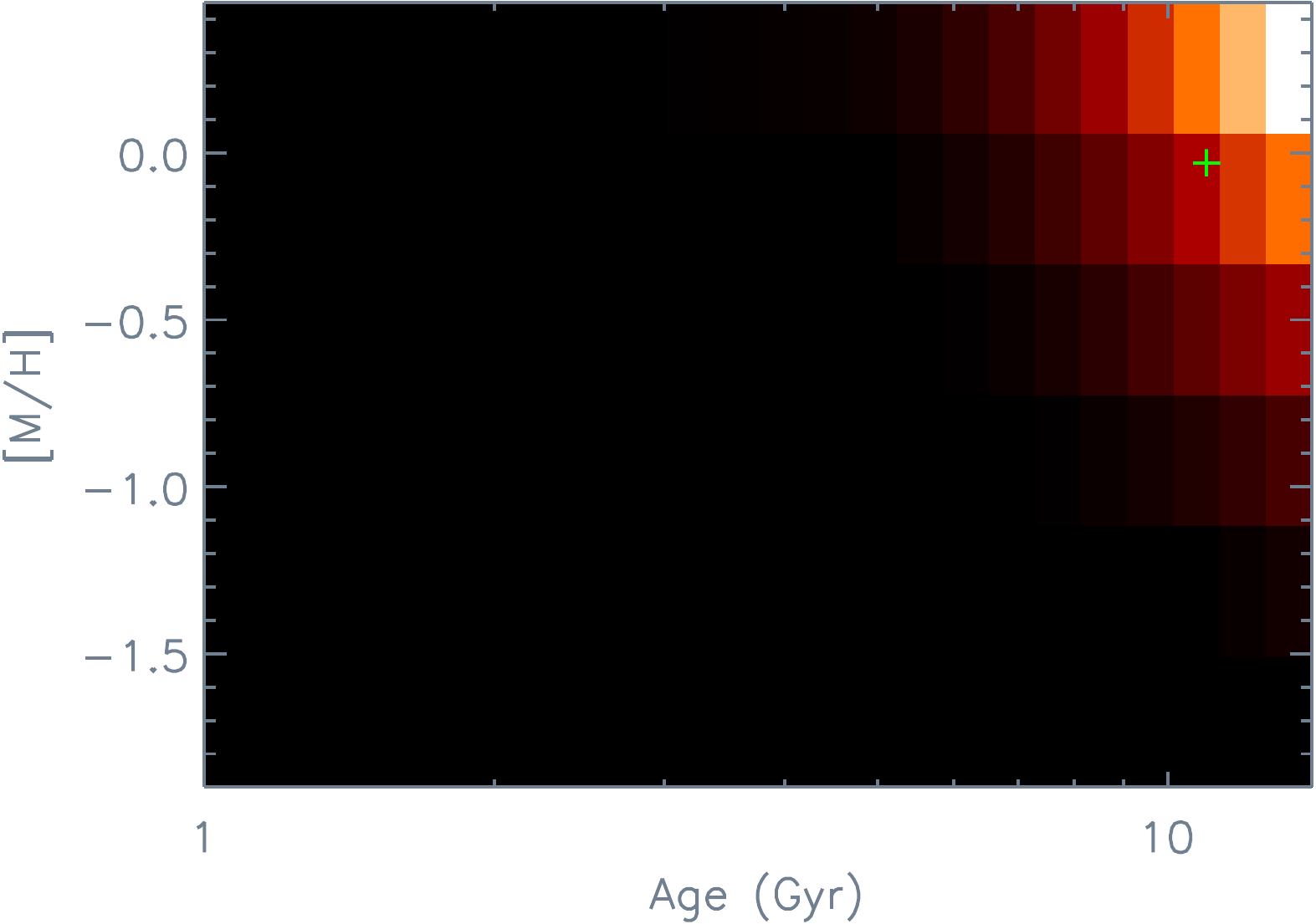}
	\end{tabular}
        \caption{{\it Top}: Magellan/IMACS spectrum of NGC~4993. The red spectrum shows the spectral fit recovered with the \textsc{gandalf} software  package. 
        {\it Bottom}: A grid containing a total of 288 stellar templates (a combination of 6 metallicities and 48 ages) used for \textsc{ppxf} fitting. Here we only plot the templates with age older than 1 Gyr. The weight of each template is represented by the strength of the color. The templates with higher weights are brighter. We present here the result for NGC~4993. The green cross represents the weighted mean stellar age and metallicity.}
        \label{host-spec_fig}
\end{figure}

We obtained an optical spectrum of NGC~4993 on 2017 September 5 (UT) using the f/4 camera of the Inamori-Magellan Areal Camera \& Spectrograph \citep[IMACS;][]{2006SPIE.6269E..0FD} on the 6.5-m Magellan/Baade telescope at Las Campanas Observatory.  We used the 600~$\ell$/mm grating with a blaze angle of $8\fdg6$ to cover the wavelength range $3500-6500$~\AA\ at a spectral resolution of $R \approx 2500$.  We obtained three 600~s exposures on NGC~4993 with a $0\farcs7$-wide long slit in mediocre conditions with some clouds.  We carried out basic reductions of the spectra (bias subtraction, wavelength calibration, flatfielding, and coaddition) using the COSMOS software package \citep{2011PASP..123..288D}.\footnote{http://code.obs.carnegiescience.edu/cosmos}  We then extracted the spectrum over a $3\farcs7$-diameter aperture in IRAF and applied a flux calibration derived from observations of the standard star LTT~6248.
The flux-calibrated spectrum of NGC~4993 is displayed in the upper panel of Fig~\ref{host-spec_fig}. 

\section{Analysis}
\label{sec:analysis}

\subsection{Stellar mass and star formation rate}
\label{host-mass}
We use the photometric redshift code \textsc{z-peg} \citep{2002A&A...386..446L}, which is based on the spectral synthesis code P\'{E}GASE.2 \citep{1997A&A...326..950F}, to estimate the host-galaxy stellar mass (\mstellar) and star-formation rate (SFR).  \textsc{z-peg} fits the observed galaxy colors with galaxy SED templates corresponding to 9 spectral types (SB, Im, Sd, Sc, Sbc, Sb, Sa, S0 and E). We assume a \citet{1955ApJ...121..161S} initial-mass function (IMF).  The photometry is corrected for foreground Milky Way reddening of $E(B-V) = 0.109$\,mag \citep{2011ApJ...737..103S,Shappee17} with $R_{V} = 3.1$ and a \citet*[][CCM]{1989ApJ...345..245C} reddening law.

\begin{figure}
	\centering
		\includegraphics*[scale=0.49]{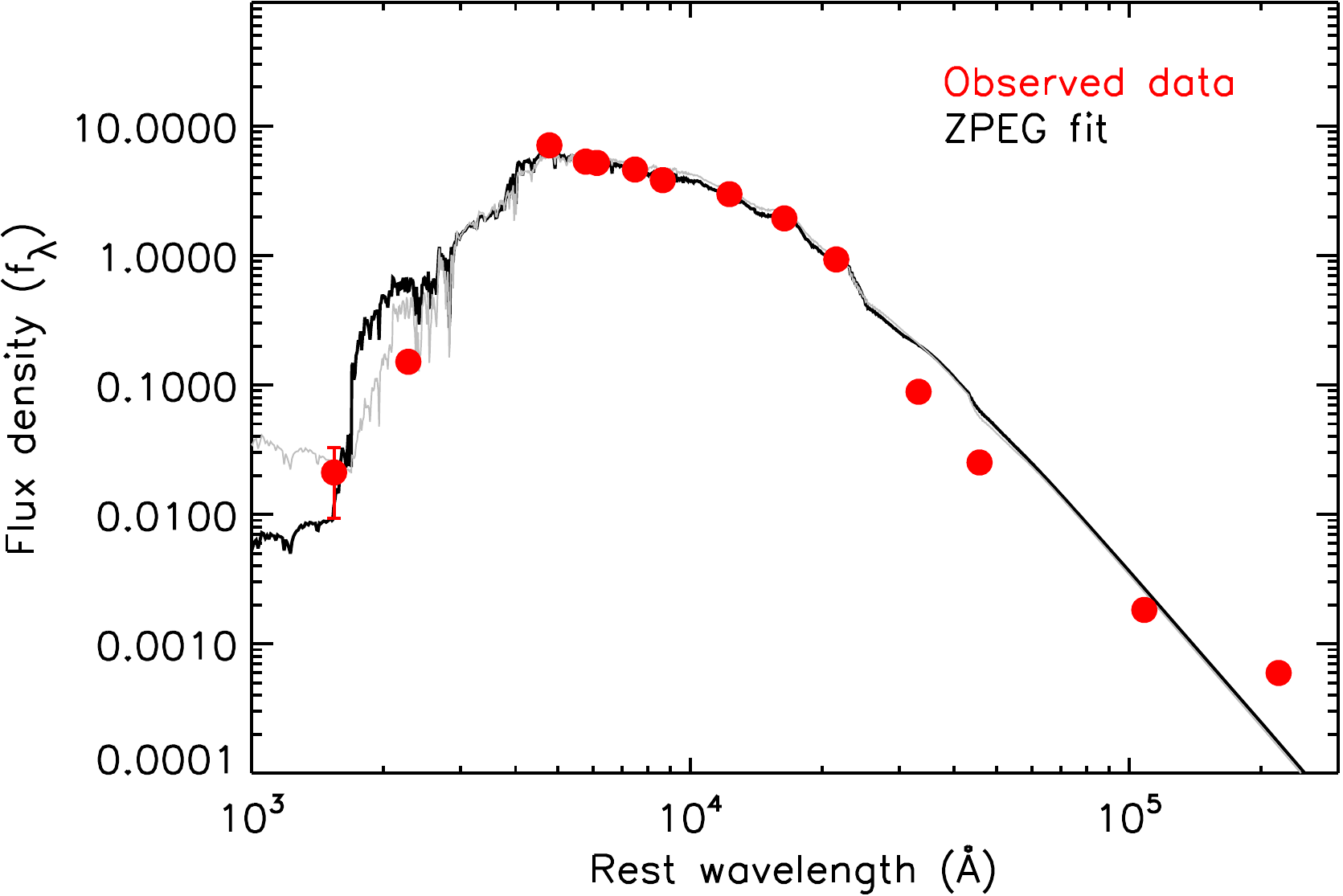}
        \caption{The best-fit SED template by \textsc{z-peg} (black curve) to the observed 14-band photometry (red filled-circle).  The grey curve represents the template by intentionally forcing \textsc{z-peg} to better fit the UV photometry but sacrificing the goodness of fitting on other bands.}
        \label{zpeg-fit}
\end{figure}

Using our 14-band photometry (see Section~\ref{sec:obs}), we measure a host \mstellar\ of $\log (M/M_{\sun}) = 10.49^{+0.08}_{-0.20}$, corresponding to a halo mass of $\mathrm{log}(M_{\mathrm{halo}}/M_{\sun})=11.96$ using the \mstellar--$M_{\mathrm{halo}}$ relation derived in \citet{2008ApJ...676..248Y}, assuming $\mathrm{log}(M_{\mathrm0}/M_{\sun})=9.8$, $\mathrm{log}(M_{\mathrm h}/M_{\sun})=10.7$, $\alpha=0.6$ and $\beta=2.9$ in their Eq.~(7).  The observed photometry and best-fit template can be found in Fig.~\ref{zpeg-fit}. 

\begin{figure*}
	\centering
		\includegraphics*[scale=0.5]{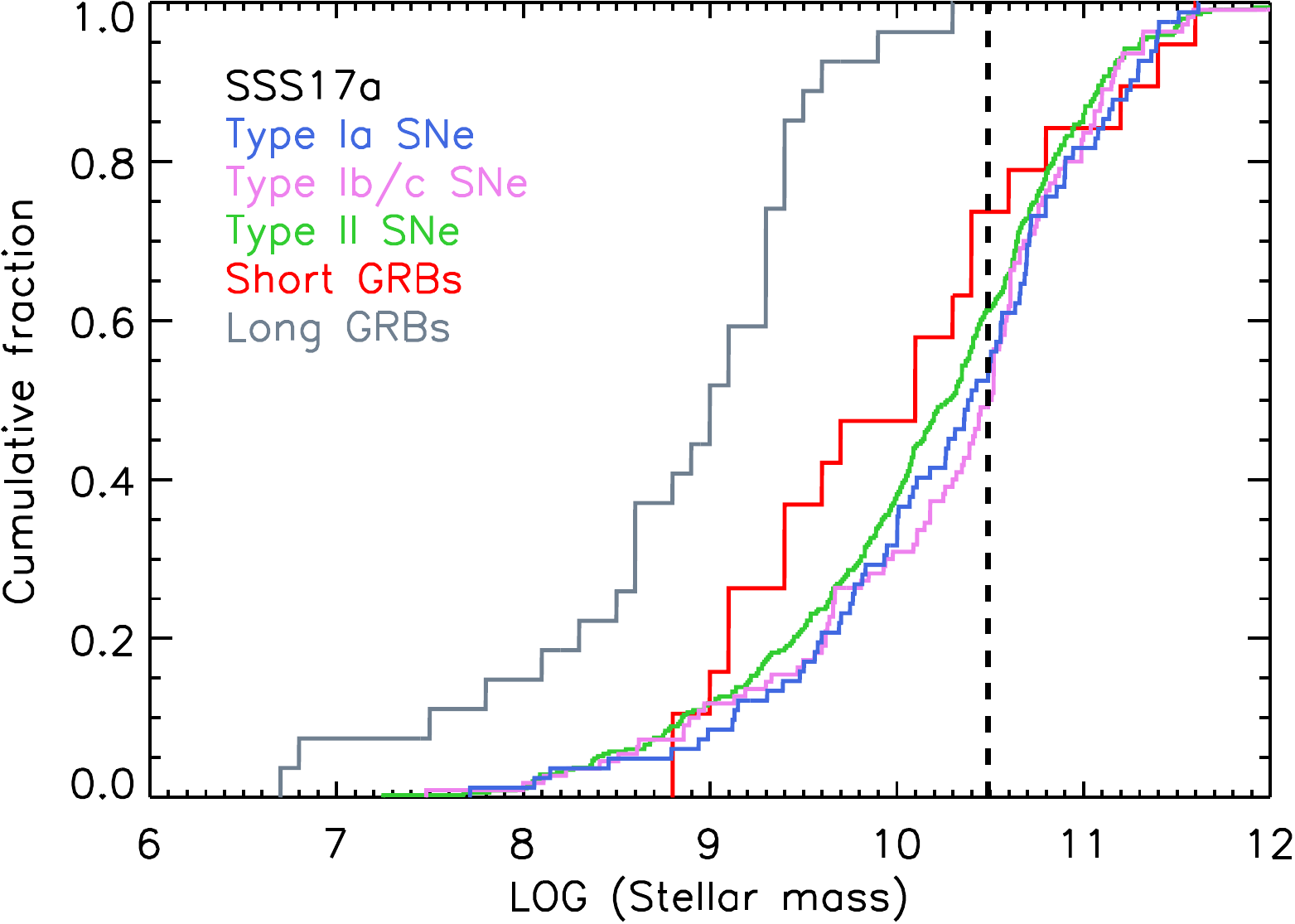}
		\includegraphics*[scale=0.5]{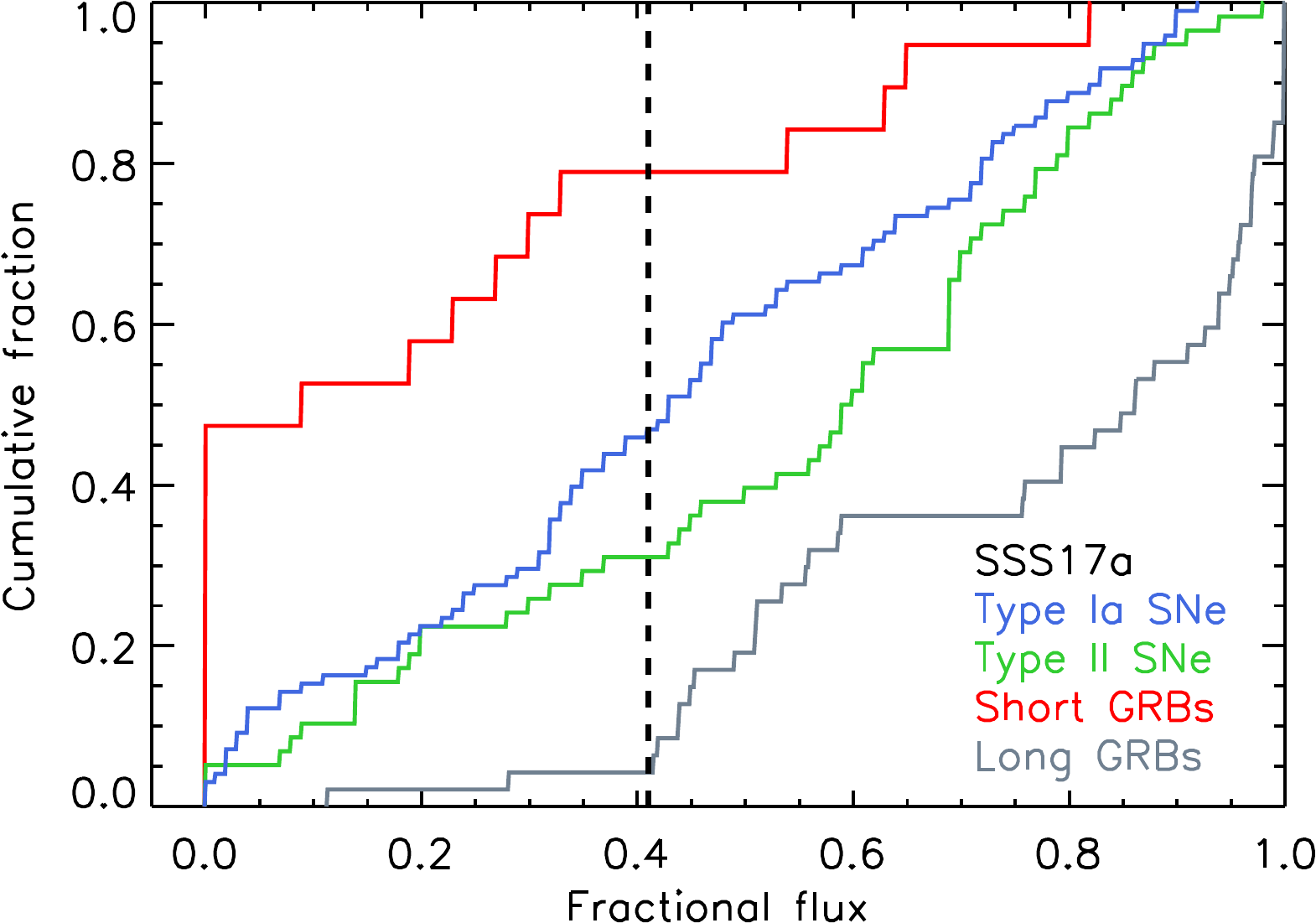}\\
		\includegraphics*[scale=0.5]{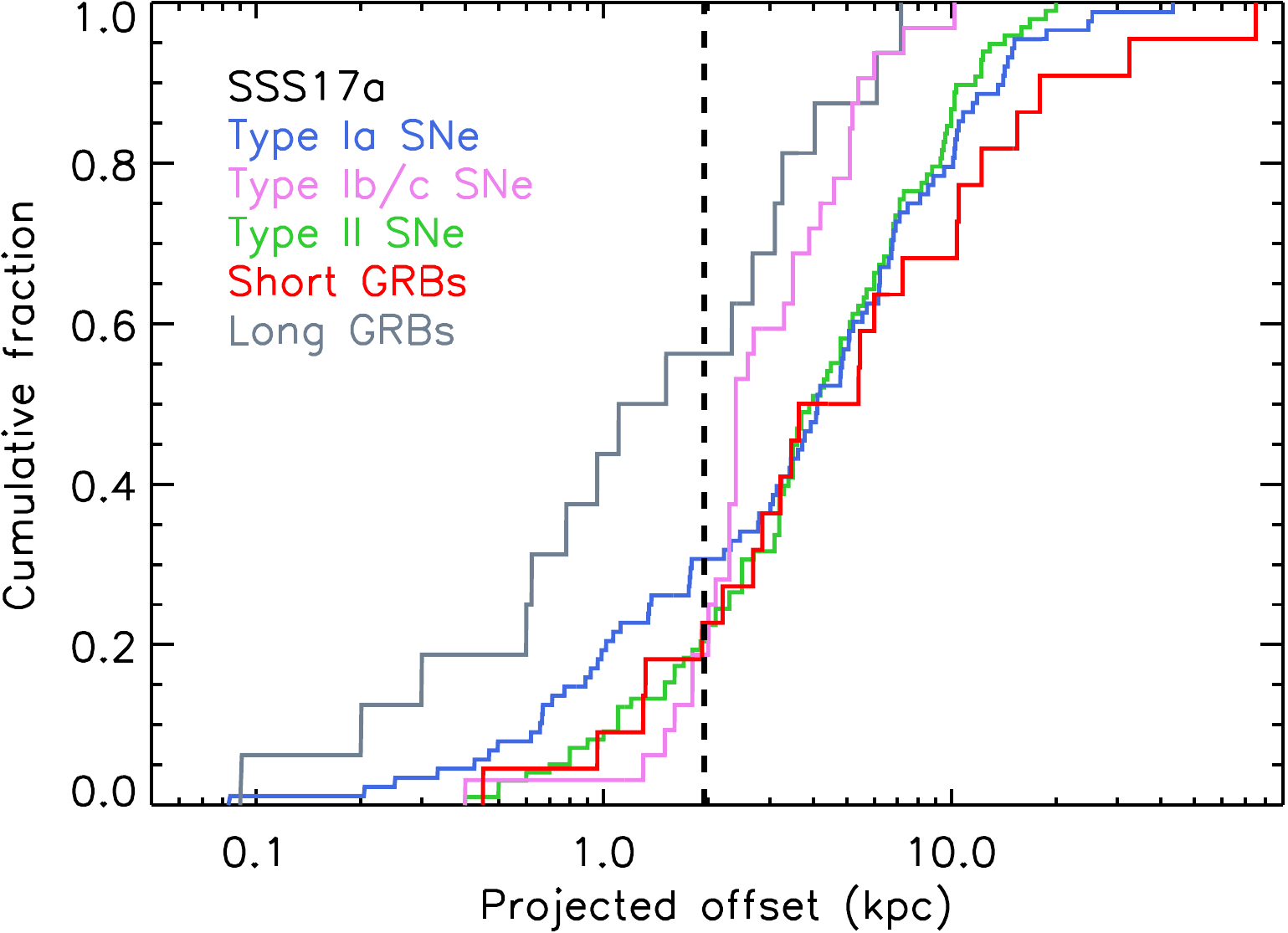}
		\includegraphics*[scale=0.5]{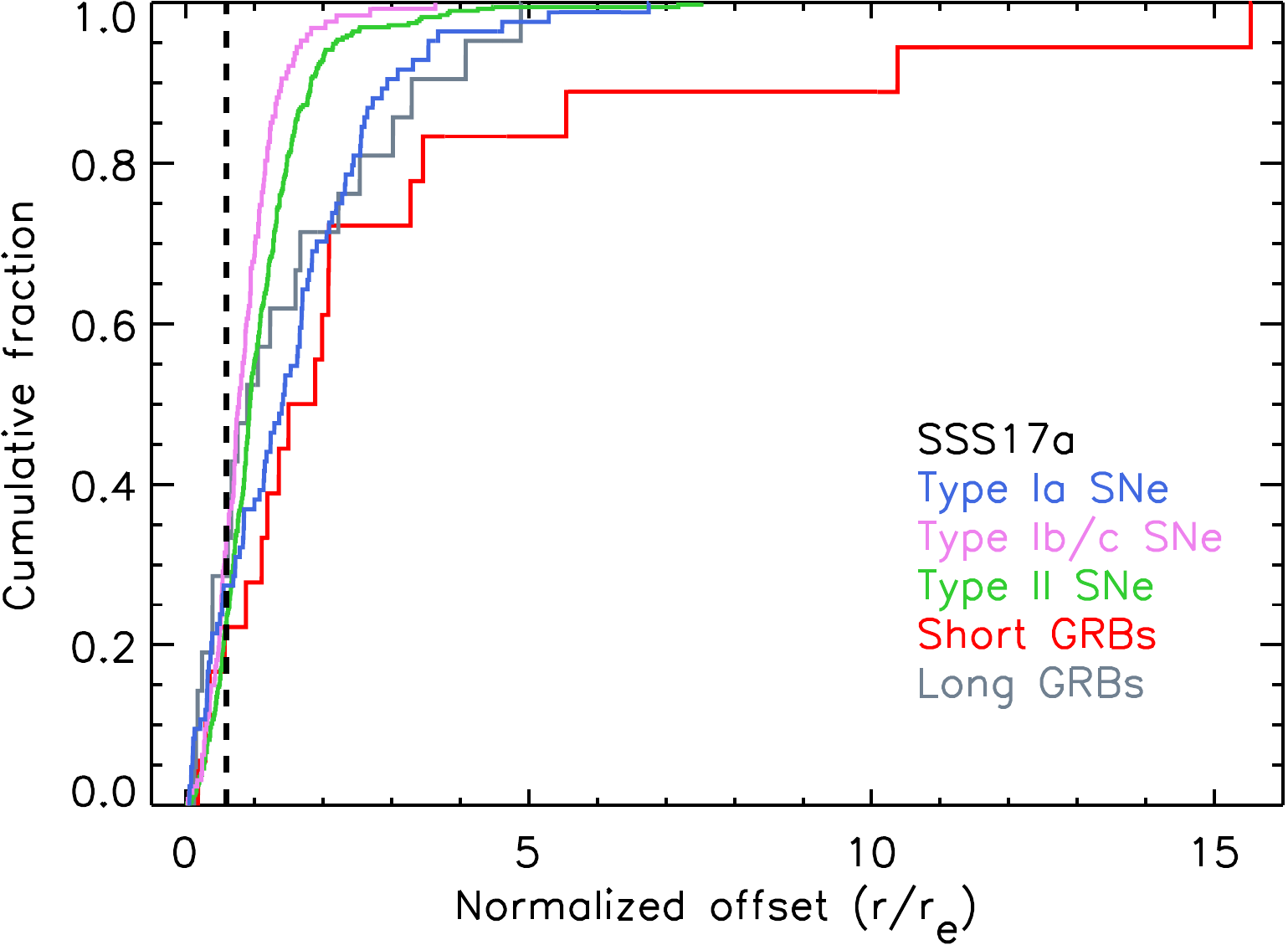}
        \caption{{\it Upper left}: Cumulative distribution of host-galaxy stellar mass for different classes of transients. The host mass of SSS17a is represented by a vertical dashed line. Also shown are the distributions for SNe~Ia \citep[blue;][]{2014MNRAS.438.1391P}, SNe~Ib/c \citep[violet;][]{2012ApJ...759..107K}, SNe~II \citep[green;][]{2012ApJ...759..107K}, sGRBs \citep[red;][]{2010ApJ...725.1202L} and long GRBs \citep[grey;][]{2010ApJ...725.1202L}.  {\it Upper right}: Same as the upper-left panel, but for the fractional flux.  Also shown are the distributions for SNe~Ia \citep{2013Sci...340..170W}, SNe~II \citep{2010MNRAS.405...57S}, sGRBs \citep{2013ApJ...769...56F} and long GRBs from \citet{2006Natur.441..463F} and \citet{2010MNRAS.405...57S}.  {\it Lower left}: Same as the upper-left panel, but for the projected offset from the host center.  Also shown are the distributions for SNe~Ia \citep{2014MNRAS.438.1391P}, SNe~Ib/c \citep{2008ApJ...673..999P}, SNe~II \citep{2008ApJ...673..999P}, sGRBs \citep{2013ApJ...769...56F} and long GRBs \citep{2002AJ....123.1111B}.  {\it Lower right}: Same as the upper-left panel, but for the normalized offset relative to the host effective radius $r_{e}$.  Also shown are the distributions for SNe~Ia \citep{2014MNRAS.438.1391P}, SNe~Ib/c \citep{2012ApJ...759..107K}, SNe~II \citep{2012ApJ...759..107K}, sGRBs \citep{2013ApJ...769...56F} and long GRBs \citep{2002AJ....123.1111B}.}\label{cdf}
\end{figure*}

In Fig.~\ref{cdf} we compare the measured \mstellar\ to that for the host galaxies of supernovae (SNe) and both short and long GRBs. Similar to SNe~Ia and core-collapse SNe, sGRBs can be found in galaxies with a wide range of \mstellar. By contrast, long GRBs are predominantly found in low-mass galaxies.  We find that NGC~4993 is more massive than 50\% of host galaxies for all classes. In fact, NGC~4993 is more massive than {\it every} long GRB host galaxy in the \citet{2010ApJ...725.1202L} sample.

\textsc{z-peg} also indicates negligible recent star formation (at least over the past 0.5\,Gyr) in the host galaxy.  The same result is obtained by intentionally forcing \textsc{z-peg} to better fit the UV photometry (but sacrificing the goodness of the full SED fitting; see the grey curve in Fig.~\ref{zpeg-fit}). This is further supported by the non-detection of nebular emission lines in the host spectrum. Using the GALEX $NUV$ photometry, we estimate a SFR of only 0.003\,M$_{\sun}$\,yr$^{-1}$ \citep[see also][]{GCN21645} based on the conversion from \citet{1998ARA&A..36..189K}. 

\subsection{Age and metallicity}
\label{age-metal}

The spectrum of NGC~4993, through its continuum and possible emission lines, provides information about its extinction, SFR, metallicity, age, and velocity dispersion.  To measure these quantities, we fit the emission lines and stellar continuum using the Interactive Data Language (\textsc{IDL}) codes \textsc{ppxf} \citep{2004PASP..116..138C} and \textsc{gandalf} \citep{2006MNRAS.366.1151S}. A complete description of this process can be found in \citet{2014MNRAS.438.1391P}. Briefly, \textsc{ppxf} fits the line-of-sight velocity distribution (LOSVD) of the stars in the galaxy in pixel space using a series of stellar templates. Before fitting the stellar continuum, the wavelengths of potential emission lines are masked to remove any possible contamination.  The stellar templates are based on the MILES empirical stellar library \citep{2006MNRAS.371..703S, 2010MNRAS.404.1639V}.  A total of 288 templates are selected with $[M/H]=-1.71$ to $+0.22$ in 6 bins and ages ranging from $0.063$ to $14.12$\,Gyr in 48 bins.

After measuring the stellar kinematics with \textsc{ppxf}, the emission lines and stellar continuum are fit by \textsc{gandalf} simultaneously. Through an iterative fitting process, \textsc{gandalf} finds the optimal combination of the stellar templates, which have already been convolved with the LOSVD. Extinction is handled using a two-component reddening model. The first component assumes a diffusive dust screen throughout the whole galaxy that affects the entire spectrum including emission lines and the stellar continuum, while the second is a local dust component around the nebular regions, and therefore affects only the emission lines. The spectral fit results from \textsc{ppxf} and \textsc{gandalf} can be found in Fig.~\ref{host-spec_fig}.

\textsc{ppxf} determines a heliocentric radial velocity $cz = 2961 \pm 5$\,\kms\ and central velocity dispersion of $161 \pm 8$\,\kms\ for NGC~4993. The best-fit value for the diffusive dust component is zero (the local dust component cannot be constrained due to the lack of nebular emissions in our spectrum), suggesting that dust extinction within the inner $3\farcs7$ of NGC~4993 is negligible.

In Fig.~\ref{host-spec_fig} we show the stellar age and metallicity distributions of the host galaxy stellar populations given by the \textsc{ppxf} fit. We determine a mass-weighted mean stellar age of 10.97\,Gyr, with the youngest and oldest stellar populations having ages of 2.8\,Gyr and a Hubble time, respectively. This result strongly suggests that the progenitor system of SSS17a was at least 2.8\,Gyr old.  Our result is consistent with previous findings that sGRBs tend to originate from older populations \citep{2010ApJ...725.1202L}.

We measure a mass-weighted mean stellar metallicity $[M/H]=-0.03$, corresponding to $\sim$0.9~$Z_{\sun}$. \citet{2010ApJ...725.1202L} used the gas-phase metallicity $\rm 12+\log(O/H)$ and measured a mean metallicity of $\sim$1~$Z_{\sun}$ for sGRB samples. They also found that the metallicities of sGRB hosts are generally higher than those for long GRB hosts (with a median metallicity of only $\sim$0.3~$Z_{\sun}$).  Therefore, NGC~4993 has a typical metallicity for an sGRB host galaxy.

\subsection{Offset and fractional flux}
\label{offset}
SSS17a is offset by $10\farcs2$ from the center of NGC~4993, corresponding to a physical (projected) offset of 1.9\,kpc using the Tully-Fisher distance of 39.5\,Mpc \citep{2001ApJ...553...47F}. In Fig.~\ref{cdf}, we compare the measured offset to that for different types of transients. It is evident that the locations of sGRBs tend to be farther from the centers of their host galaxies (with a median offset of 5\,kpc) than long GRBs and other SNe. We find that the offset of SSS17a is somewhat small in comparison to sGRBs, with $\sim$77\% of all sGRBs having an offset of $>$1.9\,kpc.  This same trend is true when normalizing the offset by the effective radius of the galaxy, where SSS17a has a normalized offset of $r/r_{e} = 0.61$, and $\sim$80\% of all sGRBs have larger normalized offsets.

To further study the local environment of the transient, we use the fractional flux method \citep[e.g.,][]{2006Natur.441..463F}. The fractional flux is defined as the sum of all flux in all pixels that are fainter than that measured at the location of the transient divided by the total flux associated with the galaxy. 
Using the {\it HST}/ACS $F606W$ image, we determine a fractional flux of 0.41 for SSS17a (Fig.~\ref{cdf}). With this metric, sGRBs do not trace the optical light of the galaxy, with $\sim$45\% of all sGRBs being at positions with effectively no galaxy light.  That is, sGRBs are often found in the far outskirts of a galaxy. In contrast, long GRBs tend to be in the brightest part of their host galaxies (with a median fractional flux of 0.86), suggesting that their progenitors are likely related to bright star-forming regions. 

The fractional flux of SSS17a is relatively high compared to sGRB samples ($\sim$80th percentile; consistent with the offset distribution), but low relative to long GRBs (only $\sim$4th percentile).

\subsection{Morphology}
\label{sec:morphology}

NGC~4993 is clearly an S0 galaxy \citep{2015A&A...581A..10C}.  To further quantify its morphology, we use \textsc{galfit} \citep{2002AJ....124..266P} to fit the surface brightness profile of NGC~4993. We fit the galaxy profile with a single S\'{e}rsic model given by
\begin{equation}
  \Sigma(r) = \Sigma_{e} \, {\rm exp} \{ -\kappa[( r/r_{e} )^{1/n} - 1] \}, \label{eqn:sersic}
\end{equation}
where $r_{e}$ is the effective radius such that half of the total flux is enclosed within $r_{e}$, $\Sigma_{e}$ is the surface brightness at the effective radius $r_{e}$, $n$ is the S\'{e}rsic index (a concentration parameter), and $\kappa$ is a variable coupled to $n$.

Fitting the {\it HST} image of NGC~4993, \textsc{galfit} gives a concentration parameter $n \approx 4$ (the de Vaucouleurs profile), which is similar to typical elliptical galaxies. The effective radius $r_{e}$ is $17\arcsec$, corresponding to a physical size of 3.3\,kpc. A residual image is created by subtracting the best-fit model from the original image (see Fig.~\ref{host-globs}).

Dust lanes are clearly seen in the residual image, extending several kpc from the galactic center (see both Fig.~\ref{host-field} and Fig.~\ref{host-globs}) roughly in the direction of SSS17a \citep{GCN21536}. However, the dust lanes do not appear to reach the position of SSS17a, providing further evidence that  SSS17a does not suffer strong extinction and consistent with the results of \citet{Shappee17}.  The dust lanes found in early-type galaxies are usually indications of recent minor mergers and likely to host active galactic nuclei \citep{2012MNRAS.423...59S}.

\subsection{Globular clusters}
\label{globs}
\begin{figure*}
	\centering
	\begin{tabular}{c}
		\includegraphics*[scale=0.38]{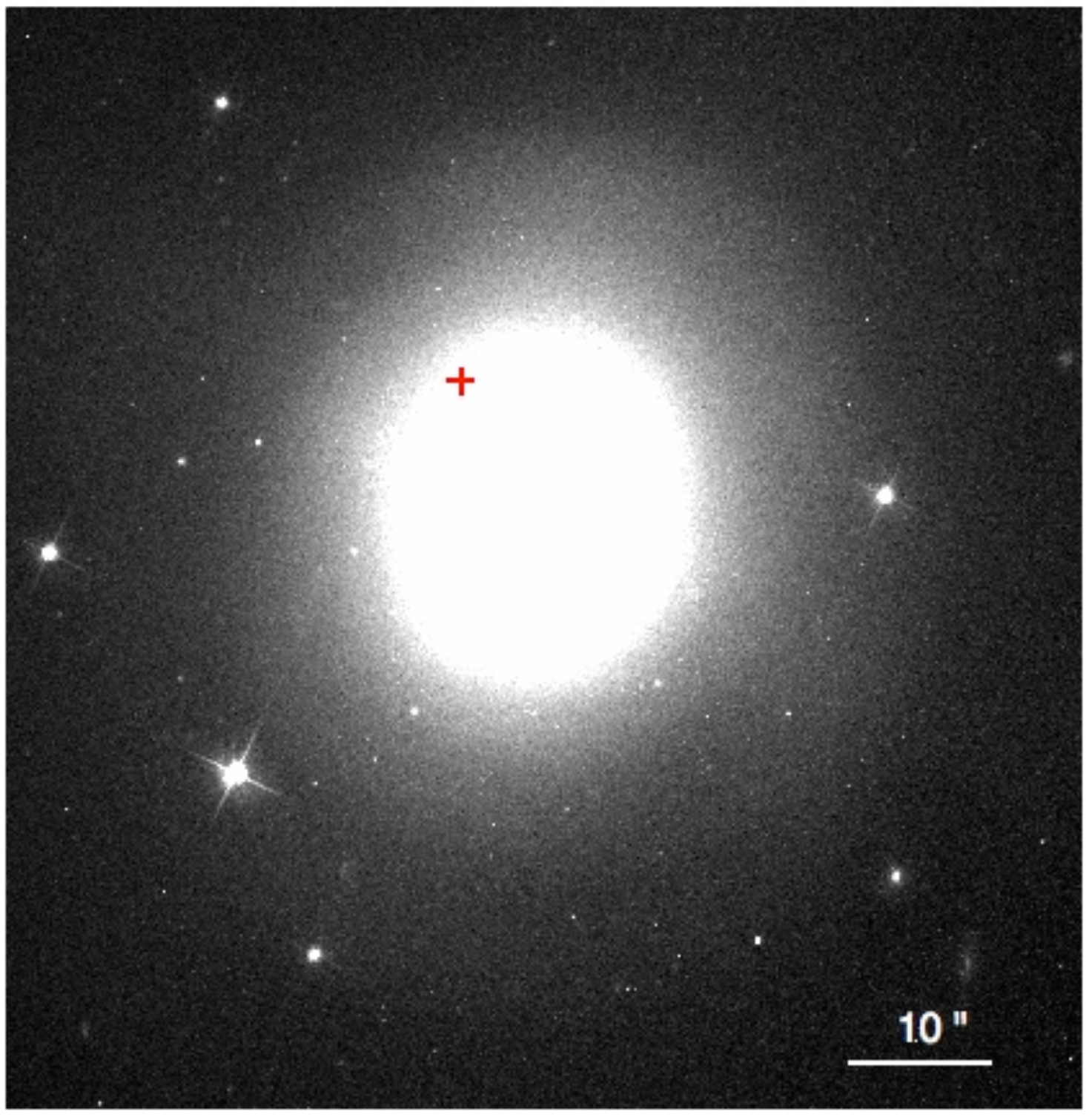}
		\includegraphics*[scale=0.38]{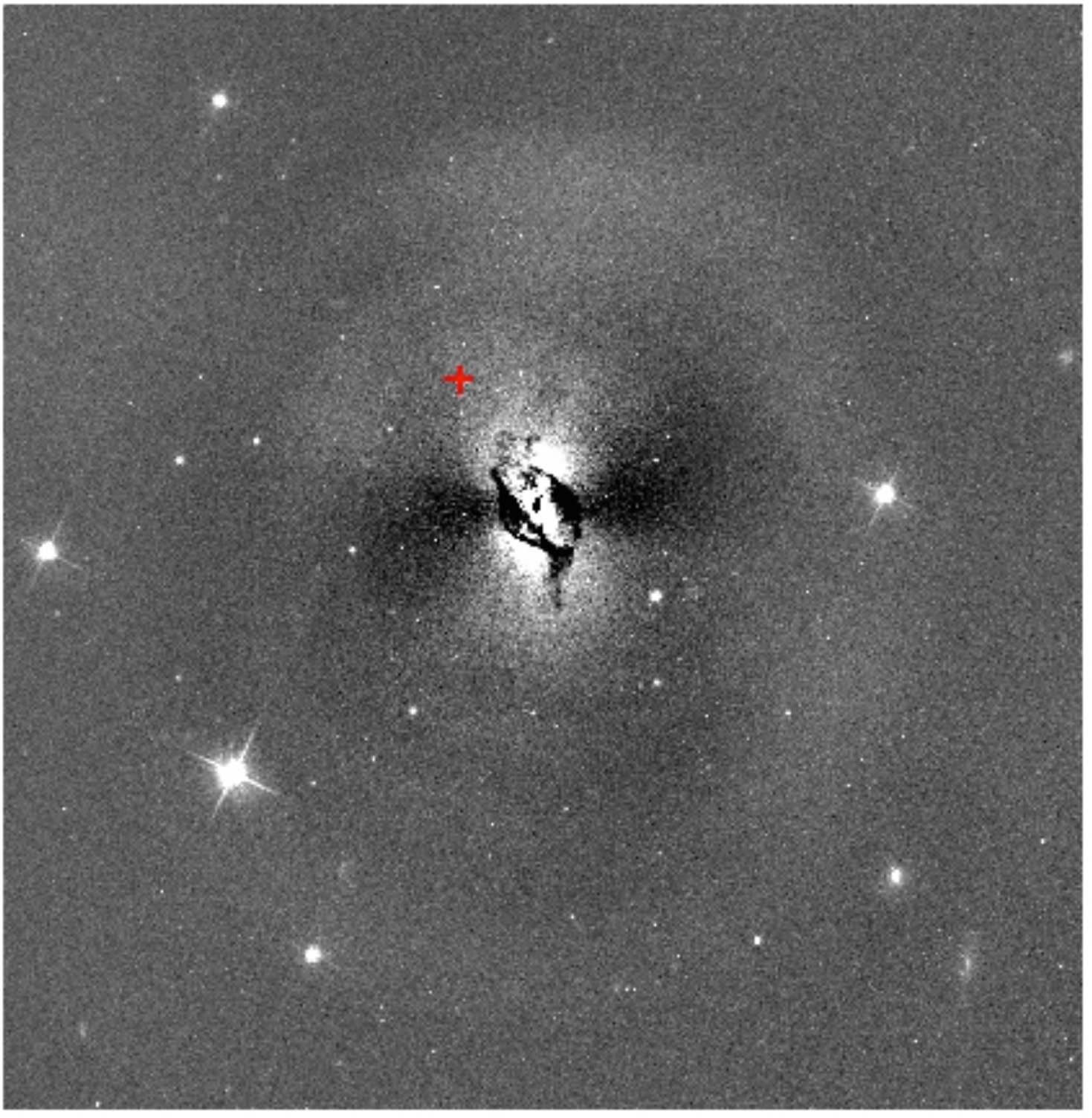}
		\includegraphics*[scale=0.38]{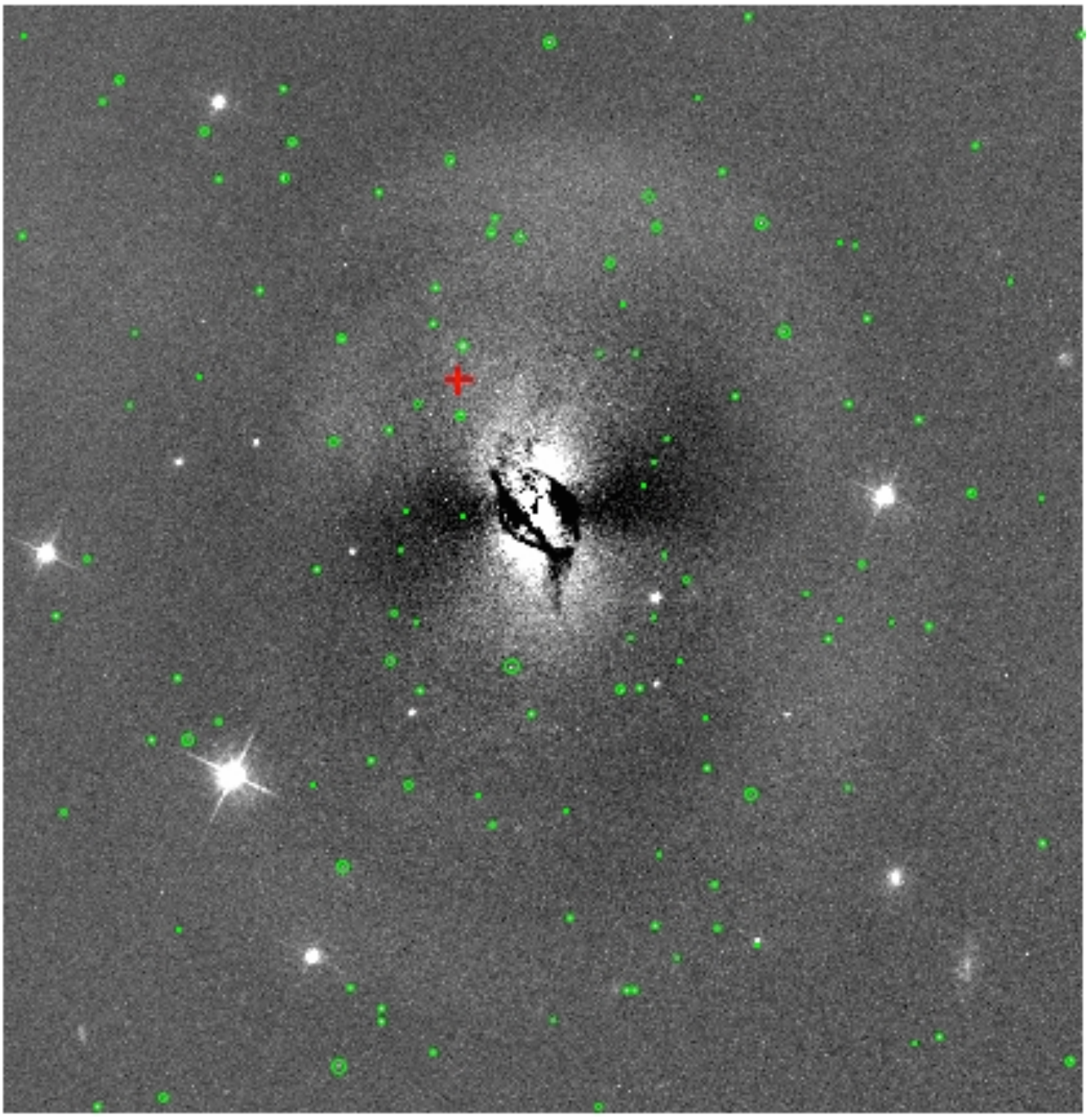}
	\end{tabular}
        \caption{{\it Left}: The {\it HST}/ACS $F606W$ image of NGC~4993. {\it Middle}: The residual image
        by subtracting a single S\'{e}rsic profile modelled by \textsc{galfit} from the same image. 
        {\it Right}: Candidate globular clusters detected (in green) after our cuts described in Section~\ref{globs}.
        The radius of the circle is proportional to size of the source. The red cross in each image
        represents the location of SSS17a.}
        \label{host-globs}
\end{figure*}

Globular clusters contain very high densities of stars. This high stellar density increases the probability of close interactions and leads to mergers more frequently than for field stars \citep{2006NatPh...2..116G,2010ApJ...720..953L,2014ApJ...784...71S}. Here we investigate the possibility that SSS17a originated from a globular cluster in NGC~4993.

To better detect sources hidden in the diffuse stellar light, we use the \textsc{galfit} residual image (Section~\ref{sec:morphology}), and identify sources using \textsc{sextractor} \citep[][see Fig.~\ref{host-globs}]{1996A&AS..117..393B}. To identify possible globular clusters, we require that each source have the following properties: (1) not obviously a foreground star (we cross-check this by using a catalog such as USNO-B1.0), (2) point-like PSF, and (3) a brightness consistent with a globular cluster at 40~Mpc given the globular cluster luminosity function \citep[e.g.,][]{2011MNRAS.416..155F}, specifically those with $21 \leq m_{AB} \leq 24$\,mag (corresponding to $-10 \leq M_{AB} \leq -7$\,mag).  A total of 119 sources pass these cuts and are selected as potential globular clusters, with the closest one being $\sim$290\,pc away in projection from the position of SSS17a. In principle, we should be able to detect all of the globular clusters in the image (the detection limit is $\sim$27\,mag). However, the number estimated here could be underestimated due to the dust extinction or the relatively bright background near the host nucleus.

Previous studies \citep[e.g.,][]{2008ApJ...681..197P} showed that the total mass of globular clusters ($M_{\mathrm{GCS}}$) within the host galaxy can be estimated by a simple scaling relation to the host galaxy halo mass ($M_{\mathrm{halo}}$) via
\begin{equation}
M_{\mathrm{GCS}}/M_{\mathrm{halo}} = \eta,
\label{eqn:gcs1}
\end{equation}
where $\eta$ represents the absolute efficiency of globular cluster formation. Assuming an efficiency $\eta\simeq4 \times 10^{-5}$ \citep{2015ApJ...806...36H} and an average globular cluster mass of $4\times10^5\,M_{\sun}$ \citep{2009MNRAS.392L...1S}, the number of globular clusters ($N_{\mathrm{GCS}}$) within a galaxy of $M_{\mathrm{halo}}$ can be estimated by

\begin{equation}
N_{\mathrm{GCS}} = (1.0 \times 10^{-10}) \times M_{\mathrm{halo}}.
\label{eqn:gcs2}
\end{equation}

Using $N_{\mathrm{GCS}} = 119$ (the number of likely globular clusters detected in the {\it HST} image), we determine $\mathrm{log}(M_{\mathrm{halo}}/M_{\sun})=12.07$, which is close to the value that we found using the \mstellar--$M_{\mathrm{halo}}$ relation (see Section~\ref{host-mass}).

\section{Discussion}
\label{sec:discussion}
In Section~\ref{offset} we show that sGRBs tend to have larger offsets from their host galaxies than other kinds of transients. The observed offset distribution is generally consistent with the predictions for compact object mergers \citep[e.g.,][]{2014ApJ...792..123B}. Simulations show that these progenitor systems experience a natal kick when the stars transition to white dwarfs, neutron stars, or black holes. The kick velocity can be up to several hundreds of kilometers per second \citep{1997ApJ...489..244F,1998ApJ...496..333F} --- potentially larger than the escape velocity of its host galaxy, which could expel the progenitor system and result in a large offset from the host galaxy.

However, SSS17a has a relatively small offset compared to the typical offsets of sGRBs. Combined with its likely old age, the location close to the center of the host galaxy suggests that the progenitor system of SSS17a was bound to NGC~4993.  Assuming a stellar mass of $\log (M/M_{\sun}) = 10.49$ (Section~\ref{host-mass}), the escape velocity of NGC~4993 is 350\,\kms\ at the transient location.  We therefore have a constraint on the SSS17a progenitor system kick of $\le350$\,\kms, which is consistent with the kicks seen for Milky Way neutron star binaries \citep{1997ApJ...489..244F,2006ApJ...639.1007W,2010ApJ...721.1689W}.

Assuming the distance to the nearest likely globular cluster (290\,pc; see Section~\ref{globs}) and the age of the youngest stellar population (2.8\,Gyr; see Section~\ref{age-metal}), a velocity of $\sim0.1$\,\kms\ is sufficient for the progenitor to travel from a globular cluster to its the current location.  Thus the progenitor kick should be dominated by the escape velocity of the globular cluster (typically several tens of kilometers per second), which makes it hard to exclude the possibility that the progenitor originated in a globular cluster.

\section{Conclusions}
\label{sec:conclusions}
In this work, we investigate the host environment of SSS17a, the first electromagnetic counterpart to a gravitational wave source. We use optical spectroscopy and broad-band UV through IR photometry of the host galaxy to constrain the host properties, such as stellar mass, SFR, age, and metallicity. Below we summarize our main findings.

\begin{enumerate}
\item[$\bullet$] NGC~4993, the host galaxy of SSS17a, is an S0 galaxy at 40\,Mpc.  It is massive and shows negligible recent star formation.  Its mean stellar age is high, suggesting that the progenitor system likely originated from an old stellar population (an age of $>$2.8\,Gyr). NGC~4993 is similar to galaxies that have hosted sGRBs and the expected host galaxies of BNS mergers.  It is unlike typical host galaxies for other transient classes, being the most distinct from long GRB host galaxies.

\item[$\bullet$] Its small projected offset combined with its likely old age suggests that the progenitor system of SSS17a was gravitationally bound to NGC~4993.  This then implies a limit on the kick velocity of the progenitor system to be $\le$350\,\kms.

\item[$\bullet$] Many likely globular clusters are detected in the host galaxy, including close to the position of SSS17a.  We cannot exclude the possibility that the progenitor of SSS17a originated from a globular cluster.

\end{enumerate}

The galactic environment of SSS17a provides additional constraints on its progenitor system beyond that extracted from the GW data and the EM observations of SSS17a itself.  With larger samples of BNS merger host galaxies, we will be able to determine if they differ in any way from sGRB host galaxies.

\section*{acknowledgments}
We thank the University of Copenhagen, DARK Cosmology Centre, and the Niels Bohr International Academy for hosting D.A.C., R.J.F., A.M.B., E.R., and M.R.S.\ during the discovery of GW170817/SSS17a.  R.J.F., A.M.B., and E.R.\ were participating in the Kavli Summer Program in Astrophysics, ``Astrophysics with gravitational wave detections.''  This program was supported by the the Kavli Foundation, Danish National Research Foundation, the Niels Bohr International Academy, and the DARK Cosmology Centre.  We would also like to thank J. Mulchaey (Carnegie Observatories director), L. Infante (Las Campanas Observatory director), and the entire Las Campanas staff for their extreme dedication, professionalism, and excitement, all of which were critical in the discovery of the first gravitational wave optical counterpart and its host galaxy as well as the observations used in this study.

The UCSC group is supported in part by NSF grant AST--1518052, the Gordon \& Betty Moore Foundation, the Heising-Simons Foundation, generous donations from many individuals through a UCSC Giving Day grant, and from fellowships from the Alfred P.\ Sloan Foundation (R.J.F), the David and Lucile Packard Foundation (R.J.F.\ and E.R.) and the Niels Bohr Professorship from the DNRF (E.R.).
A.M.B.\ acknowledges support from a UCMEXUS-CONACYT Doctoral Fellowship.
DK is supported in part by a Department of Energy (DOE) Early Career award DE-SC0008067, a DOE Office of Nuclear Physics award DE-SC0017616, and a DOE SciDAC award DE-SC0018297, and by the Director, Office of Energy Research, Office of High Energy and Nuclear Physics, Divisions of Nuclear Physics, of the U.S. Department of Energy under Contract No.DE-AC02-05CH11231.
M.R.D. and B.J.S. were partially supported by NASA through Hubble Fellowship grants HST--HF--51373.001 and HST--HF--51348.001 awarded by the Space Telescope Science Institute, which is operated by the Association of Universities for Research in Astronomy, Inc., for NASA, under contract NAS5--26555.
This research has made use of the NASA/IPAC Extragalactic Database (NED), which is operated by the Jet Propulsion Laboratory, California Institute of Technology, under contract with the National Aeronautics and Space Administration.
Based on observations made with the NASA/ESA Hubble Space Telescope, obtained from the Data Archive at the Space Telescope Science Institute, which is operated by the Association of Universities for Research in Astronomy, Inc., under NASA contract NAS 5--26555. These observations are associated with program GO--14840.

\bibliographystyle{apj}
\bibliography{sss17a_host}

\end{document}